\documentstyle[11pt,psfig]{article}
\def \etal{{\it et al.}~}
\def \kms{km s$^{-1}$}
\begin{document}

\title{Deviations from Hubble Flow in the Local Universe}

\author{Riccardo Giovanelli\\
Cornell University}

\maketitle

\section {Introduction}

Beginning with the pioneering efforts of Rubin \etal (1976), the study 
of large--scale deviations from Hubble flow has achieved undeniable 
observational successes (see reviews by Dekel 1994, Strauss and Willick
1995). In the spirit of providing fertile ground for discussion and 
further developments, this presentation will concentrate on few 
of the areas where a measure of variance of opinion exists, rather 
than on presenting a survey of results.

The determination of the peculiar velocity of galaxies over volumes
of cosmological interest is an exercise fraught with large amplitude
uncertainties. These arise from errors in the measurements, poorly
understood corrections applied to the observed parameters, ``cosmic''
sources related to the formation history of each galaxy and to its 
individual peculiarities, and statistical corrections related to the
variations in the galaxy density field. The most commonly applied
methods, which use the Tully--Fisher (1977;TF) relation for spirals and 
the $D_n$--$\sigma$ (Dressler \etal 1987) or the Fundamental Plane (FP;
Djorgovski and Davis 1987) relations for spheroidals,
lack a well developed physical basis, and an understanding of the 
impact of individual sources of scatter has been hazy at best.
It is of paramount importance that the characteristics of the error
budget of peculiar velocity measurement techniques be well understood,
in order both to correctly infer the predictive power of the method
and to derive reliable statistical corrections.
In section 2, the error budget of the TF relation will be analyzed
in detail.

The TF technique for spirals or the FP method for spheroidals requires 
a calibrated {\it template relation}. Offsets with respect to it are 
converted to peculiar velocities by means of exponential laws. It is 
well known that the estimate of peculiar velocities is independent on
the assumed value of the Hubble constant $H_\circ$, but it needs a
kinematical calibration in order for the {\it velocity zero point} to 
be established. This requires that a set of objects be found, that can 
be globally assumed to yield a TF zero--point offset, consistent 
with rest in the comoving reference frame. This is a 
toilful task, the limits of which will be tackled in section 3.

Occasionally, warning signals go off indicating that not all is 
well with the standard expectations of predictive power of peculiar 
velocity measurement techniques. For example, the peculiar velocity 
of a system as measured by using its population of spiral galaxies 
may differ from that obtained using its population of spheroidals, 
the discrepancy being well in excess of the statistical errors 
associated with each technique. It is thus legitimate to ask: are we 
comfortable that spirals and spheroidals in general yield the same 
``observed'' peculiar velocity field? We consider this question in 
section 4.

The distribution function of the cluster peculiar velocities,
which are measured with significantly better accuracy than those of
individual galaxies, has been used as a discriminant between different
cosmological models. In section 5, we review recent results on cluster
velocities that provide relatively tight constraints on models.

The early suggestion of Scaramella \etal (1989), that local motions
may maintain coherence over scales in excess of 10,000 \kms, have
received corroborating support from several sources, most notably
Lauer and Postman (1994). In section 6 we review the 
evidence for large scale bulk flows and the issue of convergence
depth in the local universe, as indicated by all--sky 
samples of relatively nearby spirals ($cz \leq 9,000$ \kms),
both in and outside clusters. 

The last few years have seen the unfolding of a massive effort in
the measurement of primary galaxy distances, with the principal aim
of reducing the uncertainty on the value of the Hubble constant.
This has also allowed a significantly improved absolute calibration 
of the zero point of the TF relation, i.e. of the ability of using
the relation to measure {\it distances}, rather than just peculiar
velocities. The availability of an increasingly accurate TF template
relation and peculiar velocity estimates to clusters such as
Virgo and Fornax allow us to test the linearity of the Hubble flow
over distances several times larger than those sampled by primary
indicators and more accurate TF estimates of the value of $H_\circ$.
We indulge in this exercise in section 7.

Finally, in Section 8 we briefly report on  ongoing efforts to improve
the definition of the local peculiar velocity field, as well as to
establish a TF template based on a set of clusters at $cz>10,000$ \kms,
which will provide a reference frame of the increased quality for all
TF work.

\section {The Error Budget of the TF Relation}

An assessment of the predictive power of the TF method and an adequate
estimate of its biases require a fair understanding of the nature 
of its associated scatter. That scatter arises from several sources:
errors in the measurements of the TF parameters and uncertainties
associated with the corrections applied to them combine with variance 
in the galactic properties produced by differences in the formation and
evolution processes. The latter is often referred to as the {\it intrinsic}
contribution to scatter; while it can in principle be reduced by flagging
objects with observable anomalies such as velocity field distortions, 
deviations from disk planarity, other gravitational and photometric 
asymmetries, etc., this approach is seldom carried out to extensive
lengths, especially when large samples of objects are studied.
Several misconceptions regarding the nature of the TF scatter appear
in the literature: that it is well represented by a single number; 
that the measurement and correction errors fully account for the
observed dispersion; that only errors on the velocity widths are important.
Giovanelli \etal (1996a; see also Rhee 1996) have made a detailed appraisal of the sources
of I band TF scatter. They find that: (i) the total TF scatter cannot
be represented by a single value; rather, it varies monotonically
with velocity width, varying by a factor of 2 between ends
of the range of widths typically used in TF applications; the average
value of the r.m.s. scatter is in the neighborhood of 0.35 mag, while
its value for the galaxies of higher velocity width approaches 0.25 mag;
(ii) measurement
and processing errors are important contributors, but they do not fully
account for the amplitude of the scatter; (iii) uncertainties on magnitude
can be important drivers of the total scatter, especially for luminous,
highly inclined galaxies; (iv) the case for an inverse TF relation that
may be assumed to be bias--free is weakened by the characteristics of
the TF scatter. These results have an important impact in the
calculation of bias corrections, particularly on the Malmquist bias
correction for objects in the far sides of sampled volumes, as well 
as in applications of the TF method for the estimate of $H_\circ$ 
(Giovanelli 1996 and \cite{Gi96a}).

\section {A TF Template Relation}

The construction of a template relation is the most delicate aspect
of a TF program. It requires careful treatment of the effect of bias,
which demands an adequate understanding of the sources of scatter,
and a kinematically satisfactory reference frame. While template
calibrations that use field galaxies have been used in the past, 
procedures that yield a template from cluster galaxy samples are 
generally preferred. The latter approach requires that: (a) the 
cluster environment does not alter the TF behavior of member galaxies, 
so that the derived template can also be applied to field objects; 
(b) bias corrections be separately estimated for each cluster;
(c) cluster motions with respect to the comoving reference frame be
measured and accounted for, before each cluster's galaxies are
incorporated in the definition of the template.
A template TF relation, shown in Figure 1 and based on a sample of 
555 galaxies in 24 clusters, with $cz$ between 1,000 and 10,000 \kms,
was produced by \cite{Gi96a}. They also verified that no environmental
effects produce detectable differences among different clusters or
between inner and outer regions of clusters.

\begin{figure} % Figure 1
\centerline{\psfig{figure=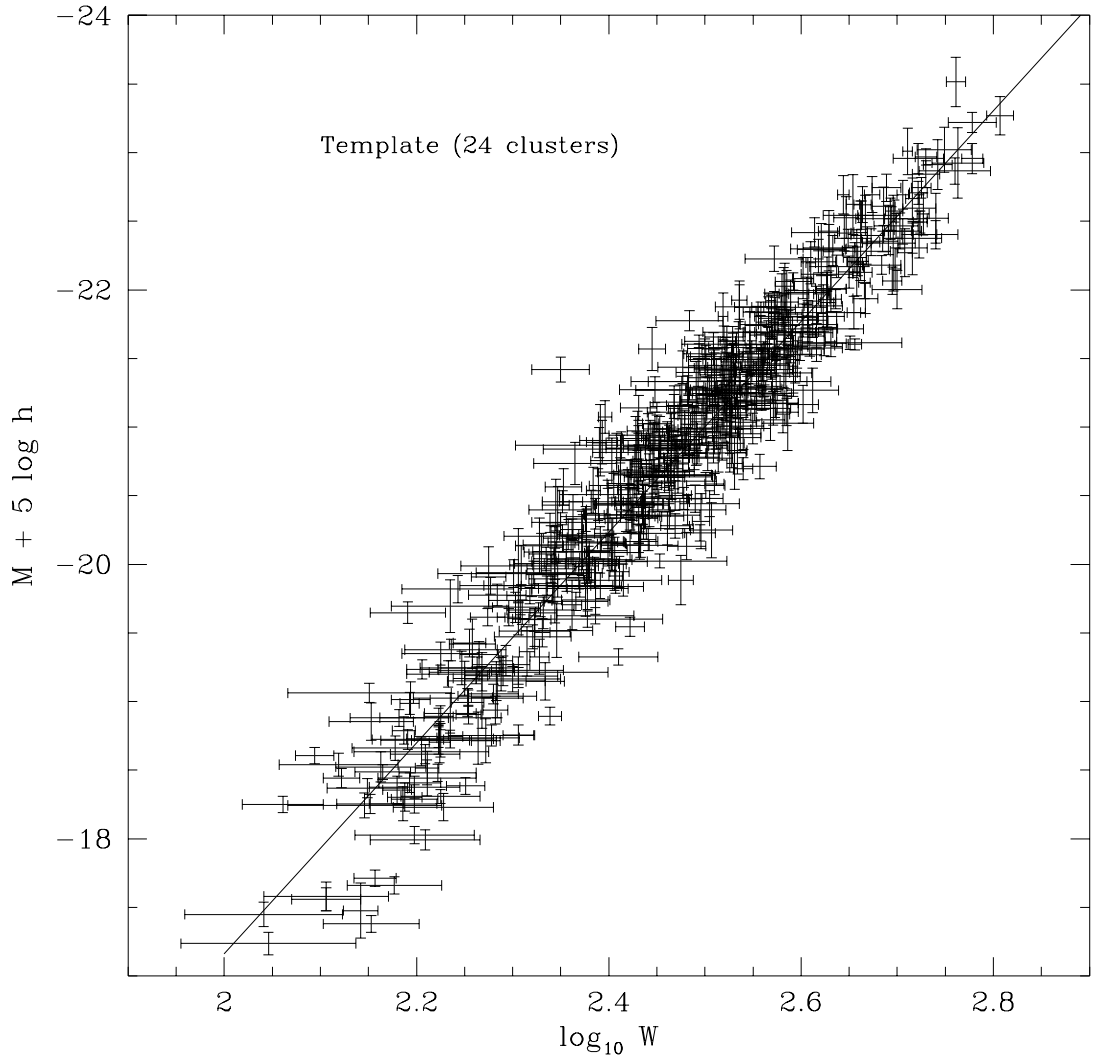,height=6.0in}}
%\epsscale{0.6}
%\plotone{fig4.ps}
\caption {Template relation based on 555 galaxies in 24 clusters 
(after [11]).}
\end{figure}

A flawed template can severely distort the observed $V_{pec}$ field.
Errors in the TF slope or offset will produce spurions $V_{pec}$,
of amplitude which will be generally variable with the target galaxy's 
velocity width. In all--sky samples, such spurious velocities
can be spotted because of their characteristic geocentric signatures.
In samples that are restricted in their sky coverage, on the other
hand, spurious modulations of the velocity field are more difficult
to identify, and they can easily be (and in some cases have been) confused
with bulk flows. The combined uncertainty on TF slope and zero point
offset, the latter constrained by sample size as well as by uncertainties
in the cluster bias corrections and the quality of the kinematical 
calibration of the cluster set, is of $\sim 0.07$ mag; this 
converts to an an uncertainty on $V_{pec}$ of $\sim 3$\% of $cz$. 

One of the principal contributors to the above mentioned uncertainty 
is that associated with the kinematical calibration. When a single
cluster is used as a reference (Coma is a frequent choice in that case),
the inferred velocity field is tied to a kinematical reference frame
where that cluster is at rest. Now, an object of radial velocity $cz$ 
and peculiar velocity $V_{pec}$ will yield a TF magnitude offset from 
the perfect template: $\delta m = 5\log (1+V_{pec}/cz)$, so that
$|\delta m|$ will decrease with increasing $cz$. If the reference
cluster $V_{pec}$ is a priori unknown, the resulting template 
will predict velocities with an associated systematic uncertainty
which, when expressed in magnitudes, will be reduced if the reference
cluster is located at higher redshift. Ideally, the kinematical reference
frame would thus be obtained using many {\it distant} clusters spread 
evenly over the sky, so that the uncertainty on the average $\delta m$
would approach zero. Let $N$ randomly distributed clusters be 
characterized by a 1--d peculiar velocity distribution function 
of r.m.s. $<V_{pec}^2>^{1/2}$ \kms; to the first order, their a priori
unknown motions will introduce a template zero point uncertainty of 
(\cite{Gi96a})

$$\Delta m \sim 2.17 <V_{pec}^2>^{1/2} <cz>^{-1} N^{-1/2} ~~{\rm mag,} 
\eqno(1)$$

\noindent
where $<cz>$ is the average cluster redshift. In \cite{Gi96a}, a subset 
of 14 clusters located between 4,000 and 10,000 \kms was used to establish 
the kinematical reference. Allowing for a measure of correlation in the
cluster distribution, estimates of $\Delta m$ vary between 0.03 and 0.06
mag. It is desirable, as we will discuss in section 8, to
extend the sample to include clusters beyond 10,000 \kms.

\section {Do Spirals and Ellipticals Yield the Same $V_{pec}$ Field?}

Peculiar velocities of galaxies have been extensively measured using
two techniques: TF for spirals and FP (or $D_n$--$\sigma$) for spheroidals.
Spiral and spheroidal samples are often combined to obtain denser 
coverage of the $V_{pec}$ field. However, unsettling indications that
the two techniques may not yield agreeable results to within the expected
level of accuracy have appeared in the literature. In Figure 2, we
plot data, adapted from Mould \etal (1991), which shows values of 
$V_{pec}$ estimated using separately spiral and spheroidal samples,
for a number of clusters; to the clusters in \cite{Mo91}, we have added
the notorious case of A2634, studied by Lucey \etal (1991).
The unflattering nature of the comparison, first pointed out by Mould 
\etal, suggests several possibilities, including 
some worrysome ones, e.g. that the assumption of universality for the 
techniques that measure $V_{pec}$ is invalid or that the impact of 
systematic errors may have been grossly underestimated.

\begin{figure} % Figure 2
\centerline{\psfig{figure=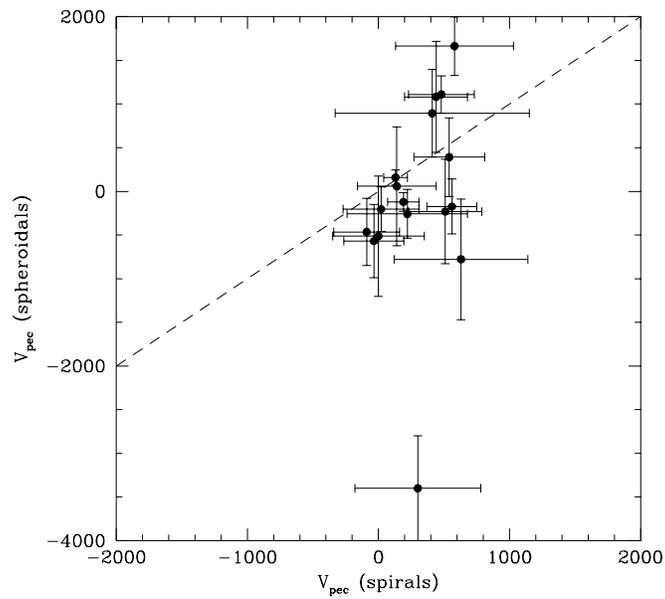,height=4.5in}}
%\epsscale{0.6}
%\plotone{fig2.ps}
\caption {Comparison of $V_{pec}$ measurements for clusters, using
spiral and spheroidal samples. The figure is adapted after Table 7
and fig. 6 of [20], with the addition of the cluster A2634,
the lowest point in the graph,
with the velocity from spheroidals of [17] and the 
velocity from spirals of [1].}
\end{figure}

Marco Scodeggio has applied his Ph.D. dissertation effort at Cornell to 
the investigation of this important issue. For each of six clusters with
large, extant TF data sets, he has obtained independent FP galaxy 
distances, carrying out a detailed comparison of the spiral and spheroidal
distances. In the particularly deviant case of A2634, a deep redshift 
survey was carried out (Scodeggio \etal 1995) in order to accurately 
characterize the structure of the cluster. It was then possible to rule 
out that the spiral and the spheroidal populations belong to separate 
dynamical entities, as well as to make accurate membership assignments. 
In A2634, new spheroidal velocity dispersions were
obtained for 55 galaxies and the TF sample of \cite{Gi96a}
was extended to include 28 cluster spirals. Scodeggio \etal (1996)
were able to verify that the new spheroidal and spiral samples yielded 
compatible values of $V_{pec}$. The comparison was corroborated in the 
case of the Coma cluster, for which 109 spheroidal and 41 spiral distances
were available. It was also verified by \cite{Sco96} and \cite{Jo96}  
that there is no significant dependence of the FP parameters 
on cluster environment. These results place much of the blame of
the problem illustrated in fig. 2 on the data quality of early
spheroidal samples. For more recent, higher quality $V_{pec}$ data, 
spiral and spheroidal galaxies appear now to yield compatible results.

\section {The Cluster Velocity Distribution Function}

The integrated, peculiar velocity distribution function of clusters of 
galaxies $P(>v)$ yields the relative number density of clusters with 
3--dimensional $V_{pec}>v$. This quantity can be estimated
for a variety of cosmological models. The observed distribution function
refers to the 1--dimensional, radial velocity component, which we here
refer to as $P(>v_{1d})$. Bahcall \etal (1994) and Moscardini \etal (1996)
have recently produced estimates of the expected $P(>v)$ for a variety
of cosmological models. Fig. 1b of Bahcall and Oh (1996) displays the 
function $P(>v_{1d})$, computed from the data of \cite{Gi96a}. 
It can be noted that:

\noindent (i) There are no clusters with peculiar velocity larger than
about 600 \kms. Given the size of our sample, one can set a conservative
limit of $P(>v_{1d}) \leq 0.05$ for $v_{1d}>600$ \kms. This is in contrast
with the values of $V_{pec}$ shown in fig. 3, for example.

\noindent (ii) Comparison of the observed $P(>v_{1d})$ with theoretical
distribution functions computed for various cosmological models is
inconsistent with high $\Omega_\circ$ models. Similar results
are obtained by \cite{Mo96}, who also find the same data 
consistent only with models with $\Omega_\circ \leq 0.4$. 

\noindent (iii) The observed 1--dimensional r.m.s. peculiar velocity of 
clusters is, according to \cite{Ba96}, $<v^2_{1d}>^{1/2} = 
293\pm28$ \kms.

\noindent (iv) Previous measurements of cluster $P(>v_{1d})$ yielded
a broader distribution, with a high velocity tail. The higher accuracy
of the measurements reported here suggests that the high velocity tail
was a spurious feature. 

\noindent (v) The cluster $P(>v_{1d})$ suggested by the density reconstruction
of Branchini \etal (1996, reported in \cite{Mo96}), has a high
velocity tail, which is in disagreement with our cluster data. A velocity
distribution with a high velocity tail, allowing for significant power
$P(>v_{1d})\geq 0.1$ at $v_{1d}\sim 1000$ \kms, would appear to be necessary 
for consistency with high $\Omega_\circ$ models.

Our cluster $P(>v_{1d})$, which suggests low $\Omega_\circ$ models,
is consistent with other cosmological cluster constraints; 
the discrepancy between the high baryon content of
clusters of galaxies and the low baryon density required by primordial
nucleosynthesis of light elements, for example, appears to be understandable
only if either the universe has low density or if the standard interpretation
of primordial nucleosynthesis is incorrect (\cite{Wh93},\cite{Lu96}).
It is however in disagreement with previous measurements of 
cluster $P(>v_{1d})$ and with derivations of the value of $\Omega_\circ$
from global mapping of the peculiar velocity field and density
field reconstructions obtained by using, for example, the Potent approach 
(\cite{De94}). These discrepancies are not yet well understood, and suggest 
that the reliability of the measured and inferred characteristics of the
large--scale peculiar velocity field still need to be considered with a 
measure of caution.

\section{Convergence Depth}
\subsection {Background}

If the number density of galaxies $\delta_N({\bf r})$ is known, and if
it can be assumed that light traces mass via a constant bias parameter $b$,
the peculiar velocity induced on the Local Group by the distribution of
masses out to distance $R$ can be written as

$${\bf V}_{pec,LG}(R) = {H_\circ \Omega^{0.6}_\circ \over 4\pi b}
\int \delta_N({\bf r}) {{\bf \hat r}\over r^2} W(r,R) d{\bf r}, \eqno (2)$$

\noindent 
where $W(r,R)$ is a window function, e.g. $W(r,R) = exp(-r^2/2R^2)$, and
${\bf \hat r}$ is the unit vector.
The asymptotic value of ${\bf V}_{pec,LG}(R)$, for $R \rightarrow \infty$
can then be matched to the velocity inferred from the CMB dipole moment
and an estimate of $\beta = \Omega^{0.6}_\circ/b$ can be obtained.
Note that the gravitational effect of a galaxy is $\propto M_{gal} r^{-2}$;
if $M_{gal}\propto L_{gal}$, then the gravitational contribution of that
galaxy to ${\bf V}_{pec,LG}$ is proportional to its flux. In principle,
${\bf V}_{pec,LG}(\infty)$ can be gauged from a catalog of positions 
and fluxes of galaxies, or any other widely distributed extragalactic
population, such as clusters. Given the discrete character of those
distributions, in practical terms eqn. (2) is replaced by a summation over 
the sampled objects, which is usually expressed in terms of the monopole 
${\cal M}(R)$ and the dipole ${\cal D}(R)$ of the distribution:

$$V_{pec,LG}(R) = (1/3)~{\beta~H_\circ R}~{{\cal D}(R)\over {\cal M}(R)}, 
\eqno (3)$$

\noindent
where

$${\cal D}(R)/{\cal M}(R) = \Bigl(3\sum {w_i \hat r_i \over r_i^2}\Bigr) 
\Bigl(\sum {w_i\over r_i^2}\Bigr)^{-1}, \eqno (4)$$

\noindent
the summations apply over all sampled objects within the distance $R$
and $w_i$ is a suitable weight, which may account for the selection function
of the catalog and the luminosity of the object. With increasing $R$, the 
monopole term tends to $4\pi {\bar N} R$, where $\bar N$ is the average
number density of objects.

Calculations of ${\bf V}_{pec,LG}(R)$ as described above have
been carried out using different catalogs, yielding diverse results.
When the predicted $V_{pec,LG}(\infty)$ are matched to the CMB dipole,
clusters of galaxies require a scaling by $\beta_c \simeq 0.3$ (Tini 
Brunozzi et al. 1995), optically selected galaxies are better fitted
by $\beta_{opt} \simeq  0.5$ (Scaramella \etal 1994) and IRAS selected
galaxies agree with even higher values of $\beta_{ir} \sim 0.9$, suggesting
varying degrees of bias between those populations and the mass distribution
(see also discussion in Strauss, this volume). 
As for the distance to which eqns. (2) and (3) need to be integrated for
the LG peculiar velocity to approximate an asymptotic value --- the
parameter usually referred to as the {\it convergence depth} ---,
different catalogs yield differing conclusions. While at least 50\%
of $V_{pec,LG}(\infty)$ seems to arise within 5000 \kms or so, the
relative importance of distant regions, i.e. at $cz>10^4$ \kms, is
still an unsettled matter. Scaramella \etal (1989) first suggested
that the Shapley Supercluster at $cz\sim 13,000$ \kms could play an
important role in affecting the LG motion. The more recent result of Lauer
and Postman (1994), confirmed by the reanalysis of Colless (1995), has 
brought added attention to the possibility that much power in the
peculiar velocity field may arise from scales in excess of $cz=10000$ \kms.

\subsection{Clusters to $cz \simeq$ 9000 \kms}

Using the template described in section 3, \cite{Gi96a} have 
obtained the $V_{pec}$ of 24 clusters and groups out to $cz \simeq 9000$ 
\kms. As discussed in section 5, their values of $V_{pec}$ are typically 
smaller than $\sim 500$ \kms, in disagreement with previous measurements 
of cluster velocities (cf. fig. 2). The
larger values of $V_{pec}$ are observed for relatively nearby groups and
for objects in the ``Great Attractor'' region.
The $V_{pec}$ field of the clusters with $4000 < cz < 9000$ 
\kms appears fairly quiescent. 
Furthermore, the dipole motion of the cluster $V_{pec}$ field,  
in the Local Group (LG) reference frame, clearly 
exhibits the ``reflex motion'' of the LG, i.e. both the dipole apex direction
and amplitude of the cluster $V_{pec}$ field mimics that of the CMB dipole:

\centerline{$V_{clu} = 577\pm 101~~$ vs. $~~V_{cmb} = 627\pm 22$}

\centerline{$l_{clu} = 272\pm 20~~$ vs. $~~l_{cmb} = 276\pm 2$}

\centerline{$b_{clu} = +31\pm 17~~$ vs. $~~b_{cmb} = +30\pm 2$,}

\noindent a good agreement, given that only 14 clusters are used for
the dipole calculation. Cluster $V_{pec}$'s have typical errors
of 2--5\% of their systemic velocity.

\subsection{Field Galaxies to $cz \simeq$ 6500 \kms}

In a collaboration referred to as ``SCI'', M. Haynes, J. Salzer, G. Wegner,
L. da Costa, W. Freudling and this author have obtained $V_{pec}$ for
a sample of $\sim 1600$ field spirals, using the TF relation with I
band CCD photometry. Their data, which extends to Dec $= -30^\circ$, has
been combined with the southern data of Mathewson \etal (1992), and 
trimmed to conform with uniform completeness criteria, to produce an 
all--sky data set which evenly samples a volume of 6,500 \kms radius. 
The data of \cite{MFB} have been reprocessed for consistency with the 
northern data, yielding a homogeneous set of $V_{pec}$ measurements.

Similarly to what is observed for clusters,  the $V_{pec}$ of field 
objects measured in the LG reference frame clearly reflects the motion 
of the LG with respect to the CMB, suggesting that the LG motion with 
respect to the CMB is largely a local phenomenon, not involving the 
majority of the volume subtended by the sample. Table 1 exhibits the 
dipole amplitudes and apices of the field spirals, segregated by redshift 
shells, and compared to the analogous quantities for the CMB dipole. It 
is clear that the outer shells of the sampled volume approach rest with 
respect to the CMB, and that the coherence length of the LG motion does 
not appear to significantly exceed the radius of the sampled volume. 
Note, however, the remarks in Strauss (this volume).

\begin{table}
\begin{center}
%\begin{minipage}{6cm}
\begin{tabular}{lrrrr}
       {$V_{cmb}$ Window}   & {$V_{pec}(LG)$}      &
       {$l$}     & {$b$} & {Nr.}   \\   
\hline
                         &            &            &           &       \\  
All                      & $417\pm28$ & $253\pm09$ & $37\pm03$ & 1585  \\
$2000-3000$ km s$^{-1}$  & $333\pm41$ & $268\pm16$ & $41\pm08$ &  235  \\
$3000-4000$ km s$^{-1}$  & $437\pm57$ & $242\pm20$ & $24\pm05$ &  303  \\
$4000-5000$ km s$^{-1}$  & $551\pm62$ & $236\pm24$ & $37\pm05$ &  372  \\
$5000-6000$ km s$^{-1}$  & $566\pm81$ & $281\pm15$ & $24\pm15$ &  270  \\
\\
CMB                      & $627\pm22$ & $276\pm03$ & $30\pm03$ &       
%\hline
\end{tabular}
%\end{minipage}
\end{center}
\caption{Dipoles of the Peculiar Velocity Field}
\end{table}

\subsection{A Test of the Lauer--Postman Bulk Flow}

The cluster and field samples described in the two preceding sections
have been combined to test the Lauer and Postman (1994; LP) bulk flow.
The latter measured the dipole of the distribution of brightest 
elliptical galaxies in 119 clusters distributed over a volume
of $\sim 15,000$ \kms radius. They found that the reference frame
defined by the group of clusters is in motion with respect to the
CMB at $689\pm178$ \kms, towards galactic coordinates 
$(343^\circ,+52^\circ)(\pm23^\circ)$. This direction is roughly 
orthogonal to the CMB apex. 

Giovanelli \etal (1996b) have selected galaxies and clusters with
measured $V_{pec}$ in their cluster and field samples, contained
within two cones of $30^\circ$ aperture, directed respectively toward
the apex and the antapex of the LP bulk flow. 
Figure 3 displays $V_{pec}$, measured in the CMB reference frame,
for galaxies and clusters within the two cones. In both cones, the
measured $V_{pec}$ fail to match the large amplitude of the 
LP flow. In Figure 4, the data displayed in fig. 3 are binned
by shells of radial velocity, and apex and antapex cones are added
in opposition, so that the sign is that of the net velocity in the
LP apex direction. The median $V_{pec}$ within a redshift of 5000 \kms
hovers between 0 and 350 \kms, while beyond that distance the component
of the flow in the LP apex direction subsides to values undistinguishable
from zero. An average bulk flow with respect to the CMB as large as
689 \kms in that direction can be excluded for galaxies within the
volume subtended by these data.

\begin{figure} % Figure 3
\centerline{\psfig{figure=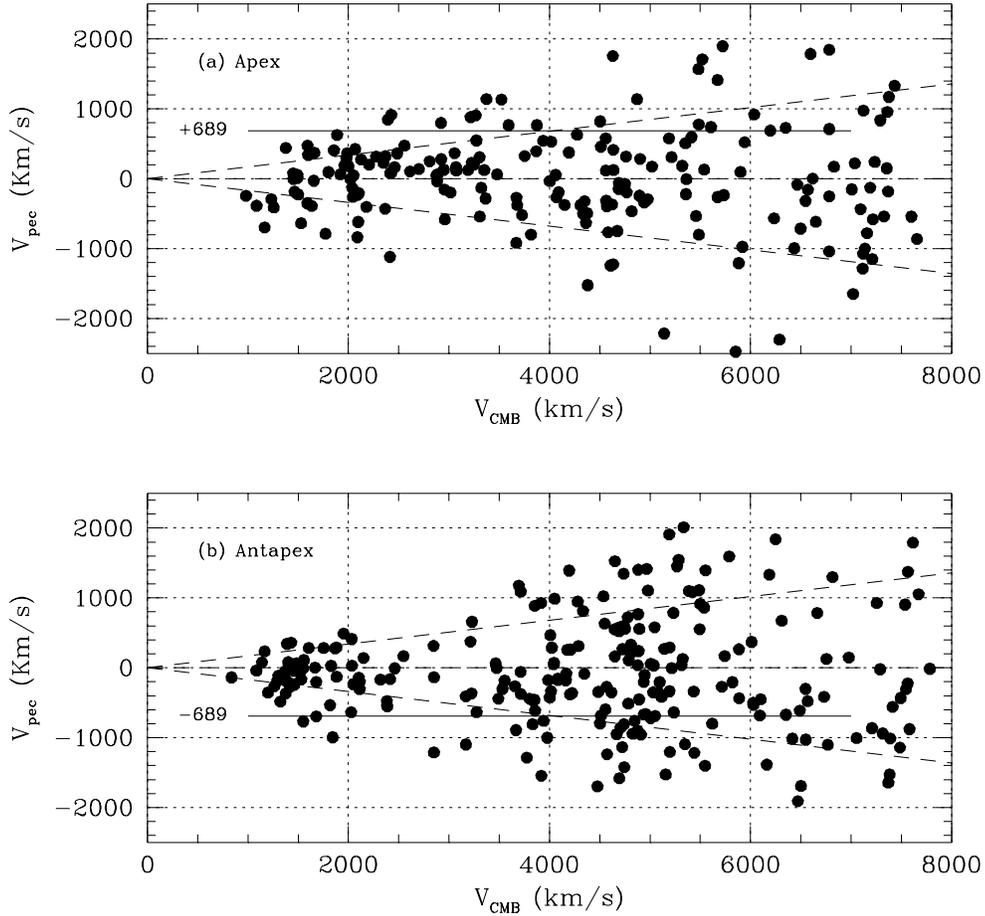,height=6in}}
%\epsscale{0.6}
%\plotone{fig2.ps}
\caption {(a) Peculiar velocities of spiral galaxies in the LP apex cone.
Both the radial velocity and the peculiar velocity are measured in the
CMB reference frame. Dashed lines refer to uncertainties deriving from a
mean TF scatter of 0.35 mag. The $+689$ \kms line refers to the amplitude
of the LP bulk flow. (b) Analogous to (a), except that it refers to the
antapex cone.}
\end{figure}

\begin{figure} % Figure 4
\centerline{\psfig{figure=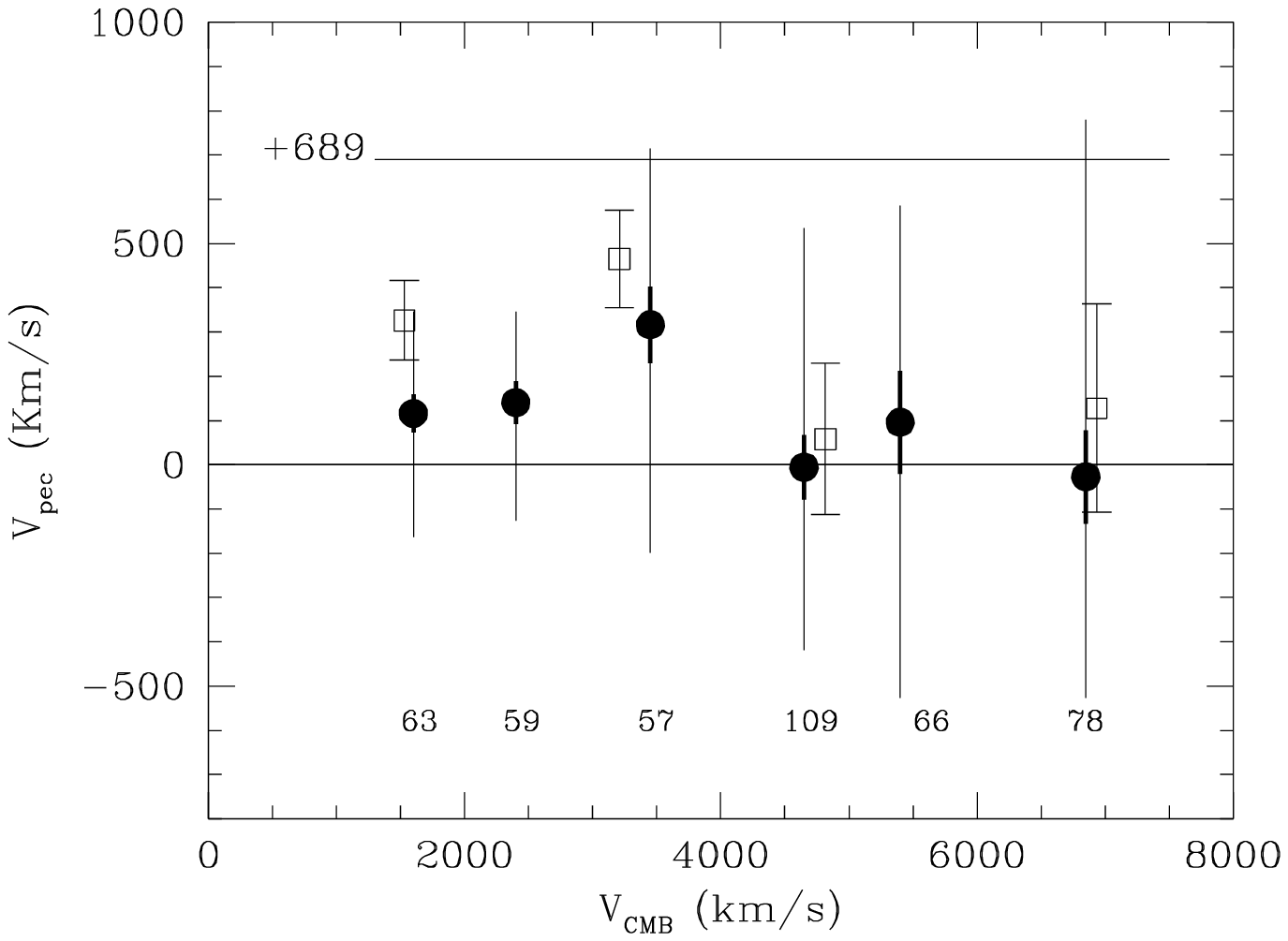,height=4in}}
\caption {Median $V_{pec}$ of field galaxies (filled symbols) and $V_{pec}$
of four clusters (unfilled symbols) in the apex--antapex cones of the
LP bulk flow. Objects in the antapex have been added in apposition to
those in the apex cone. Numbers at bottom identify the number of galaxies 
used in each bin. The $+689$ \kms horizontal line refers to the LP bulk flow 
amplitude. Thin error bars outline the range spanned by data within the 
two inner quartiles, and thich error bars approximate a standard error
on the median. The clusters included in the plot are Eridanus, ESO508,
A3574 and A400.}
\end{figure}

\subsection{A {\it Potent} Reconstruction of the Mass Density Field}

The SFI data was used by da Costa \etal (1996) to obtain a reconstruction,
shown in Figure 5, 
of the three--dimensional $V_{pec}$ and density fields, using the {\it Potent}
algorithm developed by Bertschinger and Dekel (1989); see also \cite{De94}.
The main characteristics of the reconstruction are similar to those
illustrated in recent reconstructions based on Mark III data (\cite{De94}),
differing in the following details: (i) the outline of the density field
in fig. 5 qualitatively resembles that obtained from redshift catalogs, i.e.
the mass distribution mimics that of the luminous galaxies better than
previous reconstructions; this
should help the convergence of techniques aimed at estimating $\Omega_\circ$;
(ii) the region between the Local Group and the Perseus and Coma superclusters
is dominated by voids; (iii) the Great Attractor appears to be a somewhat 
closer feature than previously found. While it is important to underscore 
the similarities between
the various density reconstructions obtained from peculiar velocity catalogs,
suggesting that real features of the density field are being identified,
it is also necessary to keep in mind that the broad window averaging required
by reconstruction algorithms yield only the coarsest features, and details
at the edges of the reconstructed maps are largely unreliable. 

\begin{figure} % Figure 5
%\centerline{\psfig{figure=fig5.ps}}
\vspace{7in}
\caption {Potent reconstruction of the density and peculiar velocity 
field using the SFI sample. See text for details.}
\end{figure}

\section{The Hubble Constant}

Recent HST measurements of Cepheid distances to galaxies in the Virgo and
Fornax clusters (see Livio \etal 1996)
have made possible better estimates of the value of $H_\circ$. 
One of the thorny issues encountered is the uncertainty on $V_{pec}$
of the clusters in which the galaxies with measured distances 
are assumed to reside. 
Expediently, estimates of the distance ratio between the given clusters 
and Coma are used, so that the uncertainty of the exercise is reduced
as the ratio $V_{pec}/cz$ for the given cluster to that of Coma.

\begin{figure} % Figure 6
\centerline{\psfig{figure=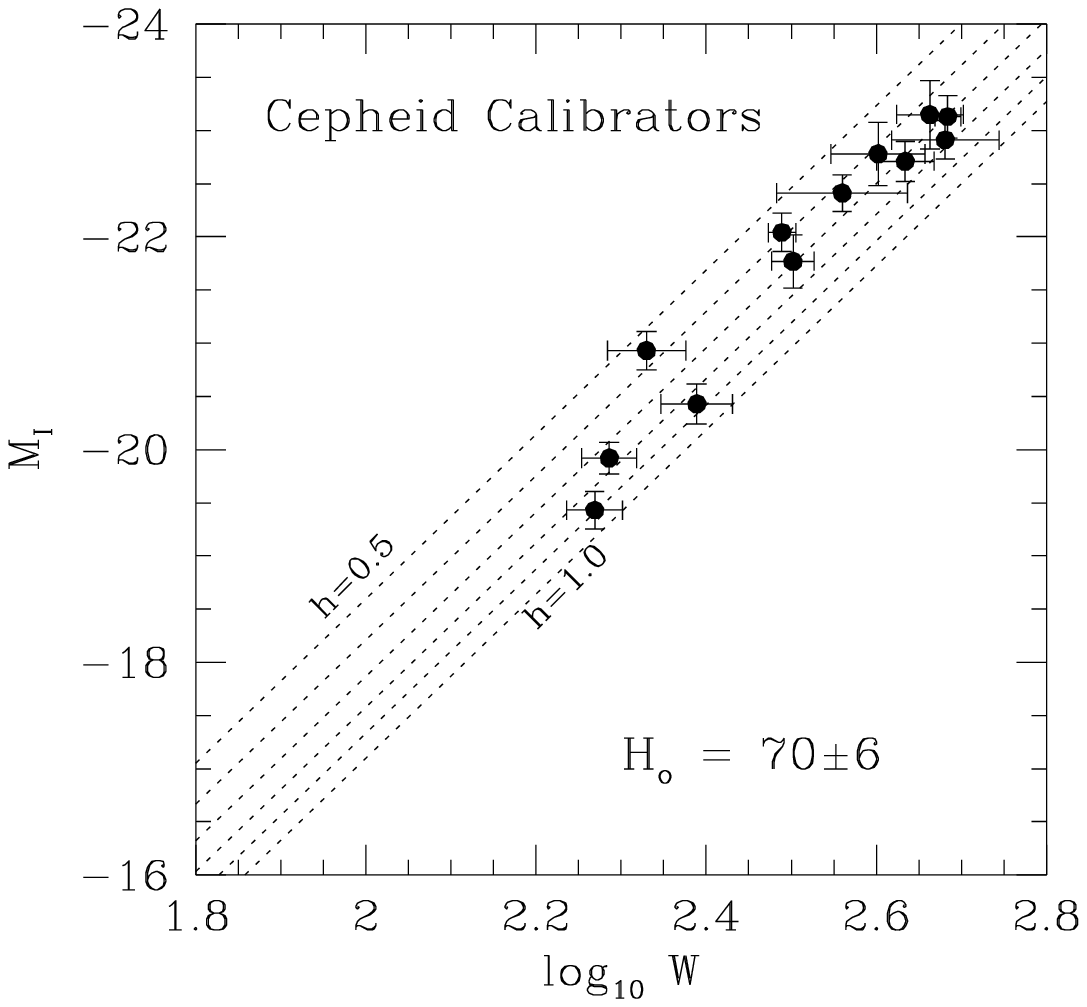}}
\caption {TF Calibrators with available Cepheid distances, superimposed
on a grid of TF template relations plotted for different values of $h$.}
\end{figure}

The {\it velocity calibration} of the TF template relation discussed above
is significantly more accurate than the ratio $V_{pec}/cz$ of any
single cluster. The accuracy of the estimate of $H_\circ$ can thus be improved 
by using the TF relation. In Figure 6, the template relation discussed in
section 3 is plotted for different values of $h=H_\circ/100$ (which is not 
constrained by the kinematical calibration of the TF relation), and the 
location of the galaxies with Cepheid distances, which are suitable for TF use.
The best match between template and Cepheid distances is for 
$H_\circ = 70 \pm 6$ \kms.
Apart from the uncertainty on the Cepheid P--L zero point calibration,
the main sources of uncertainty derive from (a) the number $N_c$ of good 
quality TF calibrators (galaxies with primary distances) and (b) the 
accuracy of the {\it velocity calibration} of the TF template relation.
As $N_c$ increases thanks to ongoing HST efforts, an improvement of (b) can
be achieved by obtaining TF distances of a number of more distant clusters 
than those currently involved in constructing our template.

\section{TF Work in Progress}

Here we refer to work in progress other than that reported in the
presentation of Strauss (this volume).
The main limitations to further applications of the TF technique to
the measurement of peculiar velocities are: (a) one of accuracy of
the method, which for relatively bright galaxies is restricted to
0.25 to 0.30 mag per single object, and (b) one of accuracy of the
parameters --- mainly the kinematical zero point --- of the TF
template relation. On the first part, explorations towards restricting
the scatter of the method by inclusion of secondary parameters, such
as color, surface brightness, etc., or by using photometry farther
in the infrared, have not produced significant improvements. Under
current circumstances, excluding low luminosity galaxies from TF
analysis appears to be the safest route to obtaining tighter relations.
Regarding the quality of the TF template, there is significant room
for improvement, which can be achieved by sampling clusters at larger
distances than those used to date. An increase of the number, of the 
regularity of the sky coverage and of the mean distance of the clusters
to be used for the definition of the TF template relation can
significantly reduce the amplitude of the systematic errors still
present in TF analysis. For field objects, dense sampling can produce
a valid description of the peculiar velocity field out to $cz\sim 10,000$
\kms. It would thus be desirable that current $V_{pec}$ surveys be
extended to that redshift.

We conclude by referring to two such extensions of the TF effort:

\noindent (1) M. Haynes and this author are currently expanding the
SCI sample to include approximately 2000 objects with available 21cm
spectroscopy, North of $-30^\circ$. The necessary I band photometry
is half--way done, and the project will be concluded in 1997.

\noindent (2) D. Dale, M. Haynes, E. Hardy, L. Campusano, M. Scodeggio 
and this author are obtaining TF data for 500 galaxies in 80 clusters
extending out to $cz\sim 20,000$ \kms. This sample will provide a
kinematical reference frame of high accuracy for the calibration of the
TF relation, as well as an accurate insight in the issue of bulk
flows on very large scales. This project is expected to be concluded
in early 1998.

\vskip 0.3in
The author wishes to acknowledge the contributions of his collaborators
in several of the results mentioned in this report, namely L. Campusano,
P. Chamaraux, L. da Costa, D. Dale, M. Haynes, T. Herter, J. Salzer,
M. Scodeggio, N. Vogt, G. Wegner. This work was supported by the National
Science Foundation grant AST94--20505.

%\section*{Bibliographic Notes} %%any end notes and further reading
                               %%suggested goes here

\vfill\eject

\end{document}